\documentclass[aps,prl,twocolumn]{revtex4}
\usepackage{epsf}
\usepackage{dcolumn}
\usepackage{bm}
\usepackage{amssymb}

\def\lsim{\mathrel{\mathstrut\smash{\ooalign{\raise2.5pt\hbox{$<$}\cr\lower2.5pt\hbox{$\sim$}}}}}
\def\gsim{\mathrel{\mathstrut\smash{\ooalign{\raise2.5pt\hbox{$>$}\cr\lower2.5pt\hbox{$\sim$}}}}}

\def\be{\begin{equation}}
\def\ee{\end{equation}}

\begin{document}

\title{Neutrino masses and the dark energy equation of state - relaxing the cosmological neutrino mass bound}

\author{Steen Hannestad}
\email{sth@phys.au.dk} \affiliation{Department of Physics and
Astronomy, University of Aarhus, Ny Munkegade, DK-8000 Aarhus C,
Denmark}

\begin{abstract}
At present cosmology provides the nominally strongest constraint
on the masses of standard model neutrinos. However, this
constraint is extremely dependent on the nature of the dark energy
component of the Universe. When the dark energy equation of state
parameter is taken as a free (but constant) parameter, the
neutrino mass bound is $\sum m_\nu \leq 1.48$ eV (95\% C.L.),
compared with $\sum m_\nu \leq 0.65$ eV (95\% C.L.) in the
standard model where the dark energy is in the form of a
cosmological constant. This has important consequences for future
experiments aimed at the direct measurement of neutrino masses. We
also discuss prospects for future cosmological measurements of
neutrino masses.
\end{abstract}


\maketitle

In the past few years a new standard model of cosmology has been
established in which most of the energy density of the Universe is
made up of a component with negative pressure, generically
referred to as dark energy. The simplest form of dark energy is
the cosmological constant, $\Lambda$, which obeys $P_\Lambda =
-\rho_{\Lambda}$. This model provides an amazingly good fit to all
observational data with relatively few free parameters and has
allowed for stringent constraints on the basic cosmological
parameters.

The precision of the data is now at a level where observations of
the cosmic microwave background (CMB), the large scale structure
(LSS) of galaxies, and type Ia supernovae can be used to probe
important aspects of particle physics such as neutrino properties.
Conversely, cosmology is now also at a level where unknowns from
the particle physics side can significantly bias estimates of
cosmological parameters.

The combination of all currently available data from neutrino
oscillation experiments suggests two important mass differences in
the neutrino mass hierarchy. The solar mass difference of $\Delta
m_{12}^2 \simeq 7 \times 10^{-5}$ eV$^2$ and the atmospheric mass
difference $\Delta m_{23}^2 \simeq 2.6 \times 10^{-3}$ eV$^2$
\cite{Aliani:2003ns}. In the simplest case where neutrino masses
are hierarchical these results suggest that $m_1 \sim 0$, $m_2
\sim \Delta m_{\rm solar}$, and $m_3 \sim \Delta m_{\rm
atmospheric}$. If the hierarchy is inverted one instead finds $m_3
\sim 0$, $m_2 \sim \Delta m_{\rm atmospheric}$, and $m_1 \sim
\Delta m_{\rm atmospheric}$. However, it is also possible that
neutrino masses are degenerate, $m_1 \sim m_2 \sim m_3 \gg \Delta
m_{\rm atmospheric}$. Since oscillation probabilities depend only
on squared mass differences, $\Delta m^2$, such experiments have
no sensitivity to the absolute value of neutrino masses, and if
the masses are degenerate oscillation experiments are not useful
for determining the absolute mass scale.

Instead, it is better to rely on kinematical probes of the
neutrino mass. Using observations of the cosmic microwave
background and the large scale structure of galaxies it has been
possible to constrain masses of standard model neutrinos. The
bound can be derived because massive neutrinos contribute to the
cosmological matter density, but they become non-relativistic so
late that any perturbation in neutrinos up to scales around the
causal horizon at matter-radiation equality is erased, i.e.\ the
kinematics of the neutrino mass influences the growth of structure
in the Universe. Quantitatively, neutrino free streaming leads to
a suppression of fluctuations on small scales relative to large by
roughly $\Delta P/P \sim - 8 \Omega_\nu/\Omega_m$
\cite{Hu:1997mj}. The density in neutrinos is related to the
number of massive neutrinos and the neutrino mass by
\begin{equation}
\Omega_\nu h^2 = \frac{\sum m_\nu}{93.2 \, {\rm eV}} = \frac{N_\nu
m_\nu}{93.2 \, {\rm eV}},
\end{equation}
if all neutrinos are assumed to have the same mass. $h$ is the
Hubble parameter in units of $100~{\rm km}~{\rm s}^{-1}~{\rm
Mpc}^{-1}$. Such an effect would be clearly visible in LSS
measurements, provided that the neutrino mass is sufficiently
large, and a likelihood analysis based on the standard
$\Lambda$CDM model with neutrino mass as an added parameter in
general provides a bound for the sum of neutrino masses of roughly
$\sum m_\nu \lesssim 0.5-1$ eV, depending on exactly which data is
used \cite{WMAP,numass}.

This should be compared to the present laboratory bound from $^3$H
beta decay found in the Mainz experiment, $m_{\nu_e} = \left(
\sum_i  |U_{ei}|^2 m_i^2\right)^{1/2} \leq 2.3$ eV \cite{mainz}.
It should also be contrasted to the claimed signal for
neutrinoless double beta decay in the Heidelberg-Moscow experiment
\cite{klapdor}, which would indicate a value of 0.1-0.9 eV for the
relevant combination of mass eigenstates, $m_{ee} = \left| \sum_j
U^2_{ej} m_{\nu_j} \right|$. Some papers claim that the
cosmological neutrino mass bound is already incompatible with this
measurement.

However, as with almost all likelihood analyses of parameters
beyond those in the simplest $\Lambda$CDM model, it is based on a
relatively limited parameter space. Using a much more complicated
model with a non-power law primordial power spectrum it is
possible to accomodate large neutrino masses, provided that Type
Ia supernova data is discarded \cite{subir}. Here, instead, we
provide a very simple (from a cosmological point of view)
extension of the standard $\Lambda$CDM model in which the dark
energy component, $X$, is represented by a fluid with a more
general equation of state, $P_X = w \rho_X$. For simplicity we
take $w$ to be a constant. Such models have been studied
extensively in the literature \cite{de} and many of them are
motivated by scalar field models, or models with modified gravity
on large scales. Most dark energy models have $w \geq -1$, but it
is possible to construct models with $w < -1$ from non-standard
kinetic terms in string theory or from models with modified
gravity \cite{modgrav}.

In fact, it is by now customary in almost all papers on
cosmological parameter fitting to allow $w<-1$. Here, we also
allow $w$ to take values below $-1$ in the so-called phantom
energy regime, and perform a standard likelihood analysis for this
model. As will be seen below, this relatively simple extension of
the $\Lambda$CDM model vastly decreases the precision with which
present observations can constrain neutrino masses.

{\it Likelihood analysis ---} We have performed a likelihood
analysis using the most recent observational data from cosmology.
We use the CMB measurements by the Wilkinson Microwave Anisotropy
Probe (WMAP) satellite \cite{WMAP}, the galaxy power spectrum
provided by the Sloan Digital Sky Survey (SDSS) collaboration
\cite{SDSS}, and the type Ia supernova data from Riess et al
\cite{SNIA}.

As parameters in the likelihood analysis we use a standard flat,
dark energy dominated model with the following free parameters:
$\Omega_{\rm CDM}$, the CDM density, $\Omega_b$, the baryon
density, $H_0$, the Hubble parameter, $n_s$, the scalar spectral
index of the primordial power spectrum, $\tau$, the optical depth
to reionization, $Q$, the normalization of the CMB spectrum, $b$,
the bias parameter, and $w = P_X/\rho_X$, the equation of state of
the dark energy. Finally we also use the contribution of
neutrinos, $\Omega_\nu h^2 = \frac{\sum m_\nu}{93.2 \, {\rm eV}}$.
From here on we assume three neutrinos with degenerate masses so
that $\Omega_\nu h^2 = \frac{3 m_\nu}{93.2 \, {\rm eV}}$. From the
flatness criterion this gives the dark energy density as $\Omega_X
= 1- \Omega_{\rm CDM} - \Omega_b - \Omega_\nu$. As additional data
we use the data from the Hubble Space Telescope (HST) Key Project
on $H_0$ which give $H_0 = 72 \pm 8 \,\, {\rm km} \, {\rm s}^{-1}
\, {\rm Mpc}^{-1}$ \cite{freedman}. We allow the bias parameter
$b$ to vary freely. While it is possible to obtain stronger bounds
on neutrino masses by adding information on the bias parameter,
this information is most likely dominated by systematics and the
errors correspondingly difficult to quantify.

\begin{figure}[htb]
\begin{center}
\epsfysize=7truecm\epsfbox{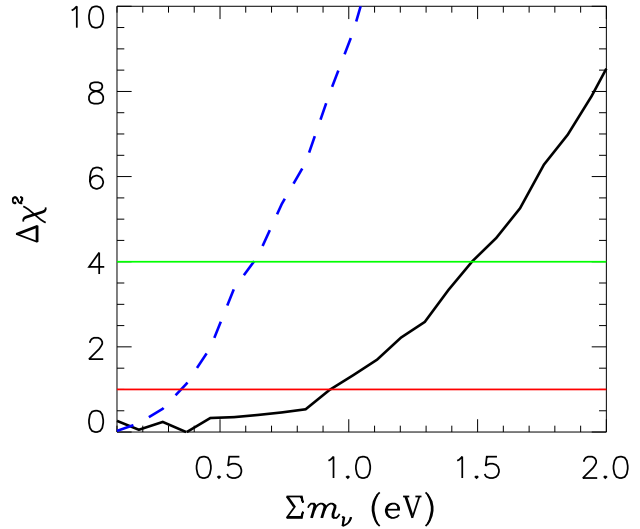}
\end{center}
\caption{$\Delta \chi^2$ as a function of neutrino mass for
various different data sets and parameter assumptions. The dashed
line is for a fixed $w=-1$, using WMAP, SDSS, HST, and SNI-a data.
The full line is for a free $w$ with the same data. The horizontal
lines show $\Delta \chi^2 = 1$ and 4, corresponding to 68\% and
95\% C.L. respectively.} \label{fig1}
\end{figure}

In Fig.~1 we show results for $\Delta \chi^2 = \chi^2 - \chi_0^2$
as a function of neutrino mass, marginalized over the other
relevant parameters (7 if $w$ is kept constant, 8 otherwise). The
marginalization was performed using a simulated annealing
algorithm \cite{SA}. The normalization constant $\chi_0^2$ is
taken to be that of the best fit model in each case. For a free
$w$ the best fit model has $\chi_0^2 = 1625.5$ for 1513 degrees of
freedom ($\chi^2/{\rm d.o.f} = 1.074$), and for fixed $w=-1$ the
best fit has $\chi_0^2 = 1626.9$ for 1514 degrees of freedom
($\chi^2/{\rm d.o.f} = 1.075$). Note that for 1-dimensional
parameter constraints we take the 95\% C.L. to be $\Delta \chi^2 =
4$. For 2-dimensional parameter estimates we take the 68\% C.L. at
$\Delta \chi^2 = 2.31$ and the 95\% C.L. at $\Delta \chi^2 =
6.17$.

From the figure, and from table I (the bottom line), it is clear
that with $w=-1$ we retrieve the known result, $\sum m_\nu \leq
0.65$ eV (95\% C.L.), published in the literature for the same
data \cite{seesaw}. However, as soon as $w$ is allowed to vary the
mass bound degrades tremendously, and in fact allows very high
neutrino masses. For the data used here the bound is $\sum m_\nu
\leq 1.48$ eV (95\% C.L.). Note that if neutrino masses are close
to saturating this bound they will be easily detectable by the
KATRIN experiment, which has a projected sensitivity of 0.2 eV for
the effective electron neutrino mass \cite{katrin}, corresponding
roughly to $\sum m_\nu \simeq 0.6$ eV. It is also completely
compatible with the claimed detection of a non-zero neutrino mass
by the Heidelberg-Moscow experiment, with a best fit around
$m_{ee} \simeq 0.1-0.9$ eV \cite{klapdor}.

\begin{table}
\caption{\label{tab:mass}The 95\% C.L. upper bound on the sum of
neutrino masses from present cosmological observations.}
\begin{ruledtabular}
\begin{tabular}{lc}
Data used & $\sum m_\nu$ \\
\hline
WMAP+SDSS+HST+SNI-a & 1.48 eV \\
WMAP+SDSS+HST+SNI-a (fixed $w$) & 0.65 eV \\
\end{tabular}
\end{ruledtabular}
\end{table}

In Fig.~2 we show the analysis in a grid for both $\sum m_\nu$ and
$w$. From this it can be seen that there is an almost perfect
degeneracy between these two parameters, an increasing $\sum
m_\nu$ can be compensated by decreasing $w$. While for low
neutrino masses a cosmological constant ($w=-1$) is allowed, for
high neutrino masses only dark energy models in the phantom regime
($w < -1$) are allowed.

\begin{figure}[htb]
\begin{center}
\epsfysize=7truecm\epsfbox{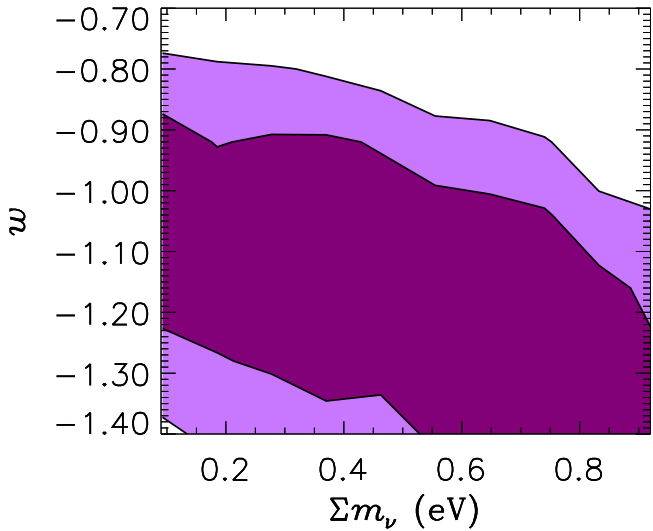}
\end{center}
\caption{68\% and 95\% allowed contours as a function of neutrino
mass and dark energy equation of state using WMAP, SDSS, HST, and
SNI-a data.} \label{fig2}
\end{figure}

The reason for the degeneracy is that when $\Omega_\nu$ is
increased, $\Omega_m$ must be increased correspondingly in order
to produce the same power spectrum. However, when $w=-1$ an
increasing $\Omega_m$ quickly becomes incompatible with the
supernova data. This can be remedied by simultaneously decreasing
$w$ because of the well-known $\Omega_m,w$ degeneracy in the
supernova data. This effect can be seen in Fig.~3: If $w$ is
allowed to vary freely, $\Omega_m$ can take very high values
without being inconsistent with the supernova data, because in
this case the $w=-1$ upper bound on $\Omega_m$ does not apply. As
soon as the model with fixed $w=-1$ approaches $\sigma m_\nu \sim
0.7$ eV and the best fit $\Omega_m$ crosses the Riess et al. bound
\cite{SNIA}, the model becomes strongly disfavoured.

\begin{figure}[htb]
\begin{center}
\epsfysize=7truecm\epsfbox{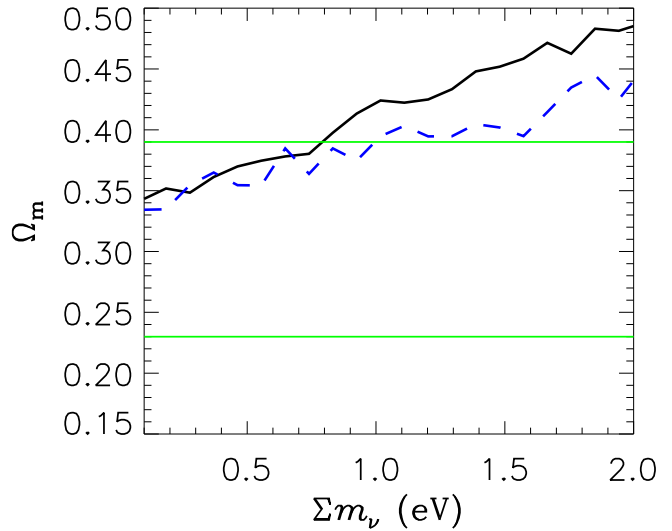}
\end{center}
\caption{The value of $\Omega_m$ for the best fit models, as a
function of $\sum m_\nu$. The curve labels are the same as in
Fig.~1. The horizontal (red) lines are the 2$\sigma$ bounds from
the present Riess et al. supernova data \protect\cite{SNIA} for
the case of $w=-1$.} \label{fig3}
\end{figure}

This model gives probably the simplest example of how to relax the
very stringent cosmological neutrino mass bound. Other means have
been discussed in the literature, such as broken scale invariance
\cite{subir} or mass-varying neutrinos \cite{massvar}, but this is
by far the simplest scenario yet discussed.

{\it Future constraints ---} Since present data clearly do not
give very stringent constraints on $\sum m_\nu$ in the presence of
phantom energy, it is worthwhile discussing whether future data
will be able to break the degeneracy. For that purpose we have
performed a Fisher matrix analysis based on a reference model with
the same free parameters as in our fit to present data
$(\Omega_{\rm CDM} h^2,w_X,\Omega_b h^2,\Omega_X,n_s,\tau,Q,b,\sum
m_\nu)$. The parameters in the reference model are
$(0.1225,-1,0.0245,0.7,1,0.05,1,1,0.05 \, {\rm eV})$. Note that we
again assume three massive neutrinos with completely degenerate
masses. While this approximation is valid as long as $\sum m_\nu
\gtrsim 0.15$ eV, it breaks down for smaller masses. This
introduces a relatively small numerical error, and does not
qualitatively change the results (see also \cite{future} for a
discussion of this point).

The Fisher matrix analysis is based on the second derivatives of
the likelihood function around the reference model and allows for
an estimate of the 1$\sigma$ error bars on measurements of all the
relevant parameters in a given experiment, assuming that errors
are Gaussian. For the future CMB data we assume an experiment
which measures both temperature and $E$-type polarization. It is
assumed to be limited by cosmic variance up to $l=2500$ for
temperature and $l=1500$ for polarization, i.e.\ an experiment
slightly better than the upcoming Planck satellite \cite{planck}
(hereafter we call this hypothetical experiment Planck+). We
neglect $B$-type polarization even though it will most likely be
measured by Planck. Adding information on $B$-type polarization
would not significantly alter the results. For a future LSS survey
we assume an effective survey volume of $V = \frac{4}{3}\pi
\lambda^3$ with $\lambda = 1000 \, h^{-1}$ Mpc (see \cite{teg2}
for a discussion). This should be compared to the SDSS-BRG
\cite{sdss2} survey which has $\lambda \simeq 620 \, h^{-1}$ Mpc.
We also assume that the linear power spectrum can be inferred with
only sampling error up to a wavenumber of $k = 0.15 h$/Mpc. In
table II we show the 1$\sigma$ error bars which can be expected on
$w$ and $\sum m_\nu$ for such a combination of data.

\begin{table}
\caption{\label{tab:error}Estimated 1$\sigma$ errors on $w$ and
$\sum m_\nu$ for the Planck + SDSS data set. The errors have been
calculated from a Fisher matrix analysis, as described in the
text.}
\begin{ruledtabular}
\begin{tabular}{lcc}
&fixed $w$ or $m_\nu$ & free $w$ and $m_\nu$ \\
\hline
$\sigma(\sum m_\nu)$ & 0.106 eV ($w$ fixed) & 0.288 eV \\
$\sigma(w)$ & 0.026  ($m_\nu$ fixed) & 0.069 \\
\end{tabular}
\end{ruledtabular}
\end{table}

From the table it is clear that if {\it either} $w$ or $\sum
m_\nu$ can be assumed to be fixed then a very stringent constraint
can be obtained on the other parameter. However, as soon as both
parameters are allowed to very freely the error bars blow up by a
factor of almost 3. The estimate that $\sigma(\sum m_\nu) \sim
0.06-0.07$ eV \cite{hierarchy}, is clearly too optimistic unless
additional information on $w$ is provided (see also \cite{future}
for further discussion). Such information could for instance come
from large scale weak lensing surveys \cite{hu05}, or from
measurements by the SNAP supernova survey satellite \cite{SNAP}.

Conversely, in order to provide a stringent constraint on $w$ it
is necessary to obtain prior knowledge about the neutrino masses.
This could come from experiments such as KATRIN \cite{katrin}
which is designed to probe the effective electron neutrino mass to
a precision of about 0.2 eV, or from neutrinoless double beta
decay experiments which in theory are sensitive to the sub-0.1 eV
range.

{\it Conclusion ---} We have studied the cosmological neutrino
mass bound in cosmological models where the dark energy equation
of state is allowed to take on an arbitrary, but constant value.
We find that this relaxes the present cosmological bound on
neutrino masses by more than a factor of two, to $\sum m_\nu \leq
1.48$ eV at 95\% C.L. Furthermore, even with the much more precise
CMB and LSS data available in the future, the degeneracy persists
unless additional data from weak lensing or similar probes can be
used to break it.

The example provided in this paper clearly illustrates that while
cosmological bounds on particle physics parameters are very
impressive, they are also model dependent. A cosmological neutrino
mass bound cannot stand completely alone, it should be
complemented by direct laboratory measurements in the 0.1 eV
sensitivity range.

{\it Acknowledgement ---} Use of the publicly available CMBFAST
package is acknowledged \cite{cmbfast}.

\end{document}